\documentclass[prl,a4paper,twocolumn,nofootinbib]{revtex4-1}
\usepackage[utf8x]{inputenc}
\usepackage{amssymb}
\usepackage{amsmath}
\usepackage{graphicx}
\usepackage{psfrag}
\usepackage{latexsym}
\usepackage{hyperref}

\newcommand{\eq}{\begin{equation}}
\newcommand{\eqx}{\end{equation}}
\newcommand{\eqn}{\begin{eqnarray}}
\newcommand{\eqnx}{\end{eqnarray}}

\newcommand{\f}[2]{\frac{#1}{#2}}

\newcommand{\nn}{{\mathcal N}}
\newcommand{\eps}{\varepsilon}
\newcommand{\teff}{T_{eff}}
\newcommand{\Lm}{\Lambda}

\newcommand{\cor}[1]{\left\langle{#1}\right\rangle}

\begin{document}

\title{The characteristics of thermalization of boost-invariant plasma from holography}

\author{Michal P. Heller}\email{m.p.heller@uva.nl}
\altaffiliation[On leave from: ]{{\it National Centre for Nuclear Research,
  Ho{\.z}a 69, 00-681 Warsaw, Poland}}
\affiliation{\it Instituut voor Theoretische Fysica, Universiteit van Amsterdam, Science Park 904, 1090 GL Amsterdam, The Netherlands}

\author{Romuald A. Janik}\email{romuald@th.if.uj.edu.pl}

\author{Przemys{\l}aw Witaszczyk}\email{bofh@th.if.uj.edu.pl}

\affiliation{Institute of Physics,
Jagiellonian University, Reymonta 4, 30-059 Krak\'ow, Poland}


\begin{abstract}
We report on the approach towards the hydrodynamic regime of boost-invariant $\nn=4$ super Yang-Mills plasma at strong coupling starting from various far-from-equilibrium states at $\tau=0$. The results are obtained through numerical solution of Einstein's equations for the dual geometries, as described
in detail in the companion article \href{http://arxiv.org/abs/arXiv:1203.0755}{\tt arXiv:1203.0755}. 
Despite the very rich far-from-equilibrium evolution, we find surprising regularities in the form of clear correlations between initial entropy and total produced entropy, as well as between initial entropy and the temperature at thermalization,
understood as the transition to a hydrodynamic description. 
For 29 different initial conditions that we consider, hydrodynamics 
turns out to be definitely applicable for proper times larger
than 0.7 in units of inverse temperature at thermalization. 
We observe a sizable anisotropy in the energy-momentum tensor at thermalization, 
which is nevertheless entirely due to hydrodynamic effects. 
This suggests that effective thermalization in heavy ion collisions may 
occur significantly earlier than true thermalization.
\end{abstract}

\maketitle

{\it Introduction.}
One of the outstanding problems of the dynamics of quark-gluon plasma (QGP) is 
the understanding of the physics of thermalization. In relativistic heavy-ion
collisions at RHIC and LHC the quantitative description of experimental data 
requires the applicability of hydrodynamics from a very early 
stage~\cite{Heinz:2004pj}. However, our insight into the non-equilibrium 
dynamics of QGP is very scarce.
The above problem is often referred to as `the early thermalization puzzle'. 
This is in fact a misnomer as viscous hydrodynamics may turn out to be applicable
when the pressures are still quite anisotropic, going against the commonly
accepted paradigm that true thermalization is necessary. One of the main results
of the present work is that for \emph{a wide range} of initial conditions this is
indeed the case. Subsequent isotropization towards true thermodynamic
equilibrium occurs purely within the quantitatively well understood
viscous hydrodynamics and is trivial in comparison. 

The key physical question of interest is the time scale after which viscous
hydrodynamic description becomes valid. This has a further refinement as viscous
hydrodynamics is really a gradient expansion with new transport coefficients 
appearing at each order. So it is very interesting to determine to what extent 
would \emph{all-order} resummed hydrodynamics describe the plasma evolution and
to what extent is one forced to incorporate genuine non-hydrodynamic degrees of freedom. 
Furthermore, the dynamics of plasma expansion will strongly depend on 
the initial state. It is very important to understand if there exists some
simple physical characterization of the initial state determining the characteristics of the transition to hydrodynamics and 
subsequent evolution. Finally, it is interesting to understand the amount of entropy produced during different stages of the dynamics.

In this letter we will address the above questions for plasma configurations
invariant under longitudinal boosts and with no dependence on transverse coordinates.
This kinematical regime was first introduced by Bjorken \cite{Bjorken:1982qr} and roughly mimicks
an infinite energy collision of infinitely large nuclei. 

Within QCD there are no techniques allowing to address these issues from first 
principles. It is thus quite natural to consider the same questions in 
the context of strongly coupled plasma in the $\nn=4$ supersymmetric gauge theory 
for the description of which
one can use the AdS/CFT correspondence \cite{Maldacena:1997re}. There, the time-dependence of plasma is translated into gravitational dynamics in 5 dimensions with a negative cosmological constant and appropriate boundary conditions. 
Using these methods perfect fluid hydrodynamics was derived at the nonlinear level
in the boost-invariant setting \cite{Janik:2005zt}, the value of shear viscosity was shown to agree \cite{Janik:2006ft}
with the one extracted from linear perturbations \cite{Policastro:2001yc}, and finally viscous hydrodynamics
was derived without any symmetry assumptions \cite{Bhattacharyya:2008jc}.

Once we consider the far-from-equilibrium regime for small proper times, gradient
or scaling expansions cease to be valid, and one has to deal with full Einstein's 
equations. Previous work by some of us \cite{Beuf:2009cx}, motivated by the early 
results of \cite{Kovchegov:2007pq}, used power series expansions around $\tau=0$ to 
study strongly non-equilibrium regime of Bjorken flow. Unfortunately, the radius
of convergence of these power series was insufficient to analyze the transition to 
hydrodynamics. On the other hand, the numerical work of \cite{Chesler:2009cy} necessarily introduced a deformation
of the physical 4-dimensional metric to pump energy and momentum into the vacuum at early times and create in this way a far-from-equilibrium state. Such a way of generating the initial state precludes the analysis of the physical evolution starting from $\tau=0$, in particular the investigation of the influence of 
the initial conditions on thermalization and entropy production that we are interested in.  
 
Motivated by this, we developed a new numerical framework using the ADM formalism of numerical
relativity and analyzed the evolution of the plasma system starting from a range of initial 
conditions. These correspond, in our setup, to specifying a single metric 
coefficient function ({\it `initial profile'}) for the initial geometry 
on the hypersurface $\tau=0$. The initial hypersurface is the same as in \cite{Beuf:2009cx}, however without any spurious coordinate
singularities. Subsequently we solve numerically 5-dimensional Einstein's 
equations and obtain plasma energy-momentum tensor from the asymptotics of the
solution at the AdS boundary. The details of this setup can be found in a companion article \cite{Heller:2012je}, while in the present 
letter we will concentrate on the physical questions mentioned above.


{\it Boost-invariant plasma and hydrodynamics.} 
The traceless and conserved energy-momentum tensor of a boost-invariant conformal plasma 
system with no transverse coordinate dependence is uniquely determined in terms 
of a single function $\cor{T_{\tau\tau}}$ -- the energy density at mid-rapidity 
$\eps(\tau)$. 
The longitudinal and transverse pressure are consequently given by
\eq
p_L=-\eps -\tau \f{d}{d\tau} \eps \quad \mathrm{and} \quad
p_T=\eps + \f{1}{2}\tau \f{d}{d\tau} \eps\,.
\eqx
It is quite convenient to eliminate explicit dependence on the number of colors $N_{c}$ and
degrees of freedom by introducing an \emph{effective} temperature $\teff$ through
\eq
\cor{T_{\tau\tau}} \equiv \eps(\tau) \equiv N_c^2 \cdot \frac{3}{8} \pi^{2} \cdot \teff^4\,.
\eqx
Let us emphasize that $\teff$ does not imply in any way thermalization. It just
measures the temperature of a thermal system with an identical energy density 
as $\eps(\tau)$.

All order viscous hydrodynamics amounts to presenting the energy-momentum tensor as
a series of terms expressed in terms of flow velocities $u^\mu$ and their 
derivatives with coefficients being proportional to appropriate powers of $\teff$, 
the proportionality constants being the transport coefficients. 
For the case of $\nn=4$ plasma, the above mentioned form of $T_{\mu\nu}$ is not
an assumption but a result of a derivation from AdS/CFT \cite{Bhattacharyya:2008jc}. 
Hydrodynamic equations are just the conservation equations $\partial_\mu T^{\mu\nu}=0$, which
are by construction \emph{first-order} differential equations for $\teff$.

In the case of boost-invariant conformal plasma this leads to a universal form of first order dynamical equations
for the scale invariant quantity
$
w=\teff\cdot \tau
$
namely
\eq
\label{eqhydro}
\f{\tau}{w} \f{d}{d\tau} w = \f{F_{hydro}(w)}{w},
\eqx
where $F_{hydro}(w)$ is completely determined in terms of the transport coefficients
of the theory, much in the spirit of \cite{Lublinsky:2007mm}. 
For $\nn=4$ plasma at strong coupling $F_{hydro}(w)/w$ is known explicitly up to
terms corresponding to $3^{rd}$ order hydrodynamics \cite{Booth:2009ct}
\eq
\label{e.third}
\f{2}{3}+ \f{1}{9\pi w} +
\f{1-\log 2}{27\pi^2 w^2}+
\f{15-2\pi^2-45\log 2+24 \log^2 2}{972 \pi^3 w^3} +\ldots
\eqx
The importance of formula (\ref{eqhydro}) lies in the fact that \emph{if} the
plasma dynamics would be governed entirely by (even resummed) hydrodynamics
including dissipative terms of arbitrarily high degree, then on a plot of 
$\f{\tau}{w} \f{d}{d\tau} w \equiv F(w)/w$ as a function of $w$ trajectories for all 
initial conditions would lie on a \emph{single} curve given by $F_{hydro}(w)/w$.
If, on the other hand, genuine non-equilibrium processes would intervene we would
observe a wide range of curves which would merge for sufficiently large $w$ when 
thermalization and transition to hydrodynamics would occur.  

\begin{figure}[t]
\includegraphics[height=4cm,width=4cm]{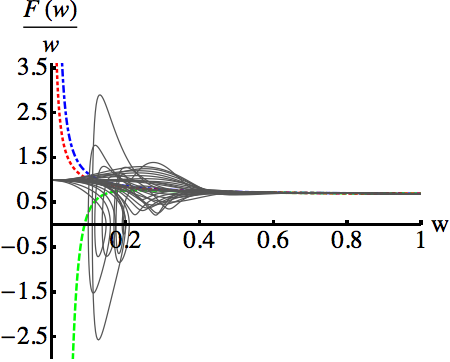}
\includegraphics[height=4cm,width=4cm]{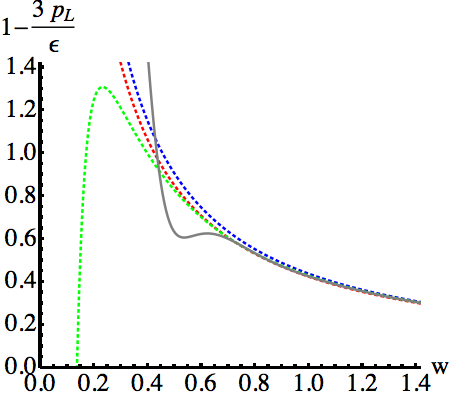}
\caption{a) $F(w)/w$ versus $w$ for all 29 initial data. b) Pressure anisotropy
$1-\f{3p_L}{\eps}$ for a selected profile. Red, blue and green curves represent $1^{st}$, $2^{nd}$ and $3^{rd}$ order hydrodynamics fit.}
\end{figure} 

In Figure~1a we present this plot for 29 trajectories corresponding to different initial states. It is clear from the plot that non-hydrodynamic modes are very important in the initial stage of plasma evolution, yet for all the sets of initial data, for $w>0.7$ the curves merge into a single
curve characteristic of hydrodynamics. In Figure~1b we show a plot of pressure 
anisotropy $1-\f{3 p_L}{\eps} \equiv 12 \f{F(w)}{w} -8$ for a selected profile 
and compare this with the corresponding curves for $1^{st}$, $2^{nd}$ and $3^{rd}$ 
order hydrodynamics. We observe, on the one hand, a perfect 
agreement with hydrodynamics for $w>0.63$ and, on the other hand, a quite sizable 
pressure anisotropy in that regime which is nevertheless completely explained by 
dissipative hydrodynamics (see \cite{Chesler:2009cy} for similar conclusion).

In order to study the transition to hydrodynamics in more detail, we will adopt
a numerical criterion for thermalization which is the deviation of 
$\tau \f{d}{d\tau} w$ from the $3^{rd}$ order hydro expression (\ref{e.third})
\eq
\label{e.criterion}
\left| \f{\tau \f{d}{d\tau} w}{F_{hydro}^{3^{rd}\, order}(w)} -1
\right| <0.005.
\eqx
Despite the bewildering variety of the
non-equilibrium evolution, we will show below that there exist, however, some
surprising regularities in the dynamics.


{\it Initial and final entropy.}
Apart from the energy-momentum tensor components, a very 
important physical property of the evolving plasma system is its entropy density $s$ 
(per transverse area and unit (spacetime) rapidity). 
In the general time-dependent case, the precise holographic dictionary 
for determining entropy is missing. 
Nevertheless in the present case due to high symmetry, 
entropy seems to be defined unambiguously in terms of $1/4 G_{N}$ of 
the apparent horizon (AH) area element mapped onto the boundary along ingoing 
radial null geodesics  
\cite{Bhattacharyya:2008xc,Figueras:2009iu,Chesler:2009cy}. 

\begin{figure}[t]
\includegraphics[height=4cm]{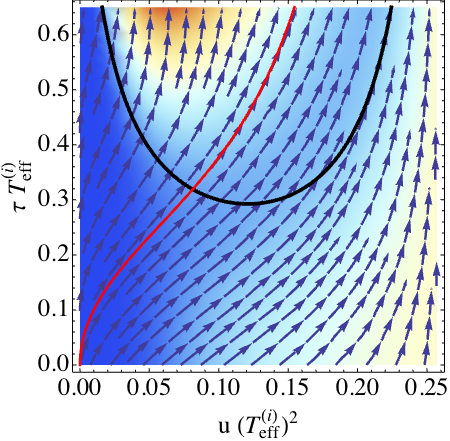}
\caption{The apparent horizon (black u-shaped curve) and a radial null geodesic 
(red curve) sent from the boundary (left
edge of the plot) at $\tau=0$ into the bulk for a sample profile.
}
\end{figure} 

For all of the initial profiles that we considered we observed an apparent 
horizon which was pierced by a radial null geodesic starting from $\tau=0$ on the boundary
(see Figure~2). This shows that the initial conditions had always some entropy 
per unit rapidity to start with. 

The main very surprising observation of our work is that the initial entropy 
density measured in units of effective temperature at $\tau = 0$ is a key 
characterization of the initial state which, to a large extent,
determines the features of the
subsequent transition to hydrodynamics as well as the final produced entropy.

In the following it is convenient to introduce a dimensionless entropy density 
$s_{n-eq}=s_{AH}/\left(\f{1}{2} N_c^2 \pi^2 (T_{eff}^{(i)})^2\right)$.
In order to evaluate the final entropy density at $\tau=\infty$, we adopted 
the following strategy. After observing a passage 
to hydrodynamics, we fitted $3^{rd}$ order hydrodynamic expression for $\teff$
\begin{eqnarray}
\teff =&& \frac{\Lm}{\left( \Lm \tau \right)^{1/3}} \Big\{ 1 - \frac{1}{6 \pi \left( \Lm \tau \right)^{2/3}} + 
\frac{-1 + \log{2}}{36 \pi^{2} \left( \Lm \tau \right)^{4/3}} + \nonumber\\
&&+ \frac{-21 + 2\pi^{2} + 51 \log{2} - 24 \log^{2}{2}}{1944 \pi^{3} \left(\Lm \tau\right)^{2}}
\Big\}
\end{eqnarray}
to obtain the remaining single scale $\Lm$. Since at $\tau=\infty$ 
perfect fluid hydrodynamics applies, we can use the standard expression for
entropy to get $s_{n-eq}^{(f)}= \Lm^{2}  \cdot \big(T_{eff}^{(i)}\big)^{-2}$.

\begin{figure}[t]
\includegraphics[height=2.6cm]{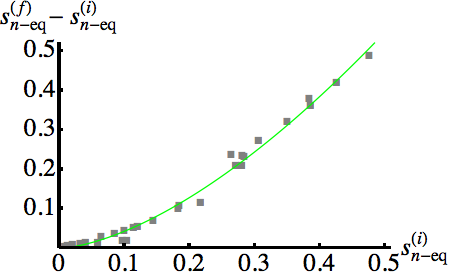}
\caption{Entropy production as a function of initial entropy.
}
\end{figure} 

Once this has been done we can now determine the entropy production 
$s_{n-eq}^{(f)}-s_{n-eq}^{(i)}$ as a function of $s_{n-eq}^{(i)}$ for all the considered 
profiles. Despite the huge differences in the evolution evident in Figure~1a, 
we observe a clear functional dependence of the entropy production on
the initial entropy. The results are shown in Figure~3 together with a fit
$
s_{n-eq}^{(f)}-s_{n-eq}^{(i)} = 1.64 \cdot \big(s_{n-eq}^{(i)}\big)^{1.58}.
$

{\it Properties of thermalization.}
We will now proceed to study in detail the properties of the transition from
far-from-equilibrium regime to hydrodynamics. We will adopt the criterion 
(\ref{e.criterion}), which imposes quite precise agreement between the equations
of motion coming from third order hydrodynamics (being the most precise description currently known) and the actual evolution of the energy density of the plasma 
obtained from numerically solving the full Einstein's equations. This criterion is quite different from criterions based on 
isotropization of the longitudinal and transverse pressures like the one adopted 
in \cite{Beuf:2009cx}. In particular, Figure~1b shows quite a sizable pressure anisotropy, which is nevertheless entirely due to hydrodynamic modes. 

Using the criterion (\ref{e.criterion}), we determine the thermalization times 
for 29 initial profiles. If we were to modify the threshold, the thermalization 
time would of course shift but for most profiles not significantly \cite{Heller:2012je}. 
However,
it is fair to say that thermalization is not a clear-cut event but rather 
happens in some narrow range of proper times.

With this proviso we will now proceed to analyze the following features of the 
thermalization time: (i) the dimensionless parameter $w=\teff \cdot \tau$, 
(ii) the thermalization time in units of initial temperature and (iii) the 
ratio of the effective temperature at the time of thermalization to the initial 
(effective) temperature.

In Figure~4, we show a plot of the values of $w$ at the time of thermalization as
a function of the initial entropy. We see that for a wide range of initial
entropies, the values of $w$ at thermalization are approximately constant and 
decrease only for initial data with very small entropies.

\begin{figure}[t]
\includegraphics[height=2.6cm,width=4cm]{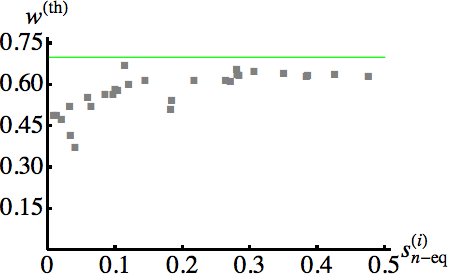}
\caption{The dimensionless parameter $w=\tau \, T_{eff}$ at thermalization 
as a function of the initial entropy $s_{n-eq}^{(i)}$. The straight line 
corresponds to $w^{(th)} = 0.7$.}
\end{figure} 

\begin{figure}
\includegraphics[height=2.5cm,width=4.0cm]{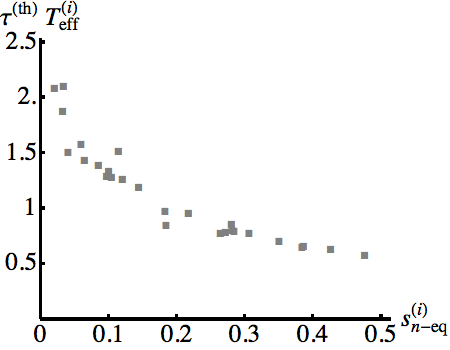}
\includegraphics[height=2.5cm,width=4.0cm]{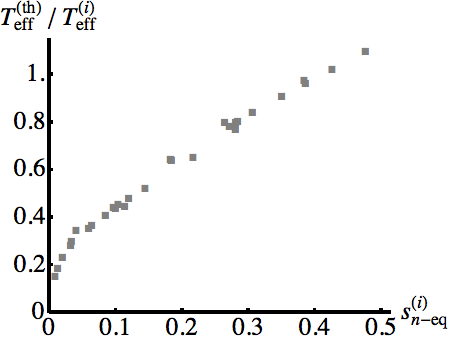}
\caption{The thermalization time in the units of initial effective temperature 
(left) and the ratio of thermalization and initial effective temperatures $T_{eff}^{(th)}/T_{eff}^{(i)}$ (right) as functions of the initial entropy $s_{n-eq}^{(i)}$.}
\end{figure} 

Subsequently, we found an unexpectedly strong correlation of the thermalization time with the initial entropy (see Figure~5). 
This is very surprising taking
into account the huge qualitative differences in the evolution of the plasma
when starting from various initial conditions.

Another important aspect is the question which part of the cooling process 
of the plasma occurs in the far-from-equilibrium regime and 
which part occurs within hydrodynamic evolution. This can be quantified by the
ratio of the effective temperatures at thermalization time and at $\tau=0$. 
The plot in Figure~5 shows very clear correlation of this quantity with 
the initial entropy. The meaning of the points with high entropy requires some
comment. We found that for these initial conditions, the energy density initially
rises and only later decreases, thus even a ratio of $T_{eff}^{(th)}/T_{eff}^{(i)}$ close 
to 1 is realized after a sizable non-equilibrium evolution \cite{Heller:2012je}.


{\it Conclusions.}
The crucial new feature of the holographic studies of Bjorken flow reported here is 
the ability to track physical observables from the far-from-equilibrium regime
at $\tau = 0$ up to thermalization and subsequent hydrodynamic evolution
without introducing any deformations in the field theory lagrangian. 
The initial state is highly anisotropic, in particular always has a negative 
longitudinal pressure \cite{Kovchegov:2007pq,Beuf:2009cx}. 
Despite the very rich early time dynamics, which depending on the initial state might 
have a plateau, a bump or a sharp decrease in the effective temperature as a function 
of proper time, we uncovered surprising regularities in the behavior of total produced 
entropy and effective temperature at thermalization as functions of initial entropy 
(all measured in units of effective temperature at $\tau = 0$). 
An interesting curiosity is that despite describing an expanding medium, the effective 
temperature at thermalization might be higher than the initial one for initial states 
with sufficiently big entropy.
For initial states with small entropy, the energy density at thermalization is much 
smaller than the one at $\tau=0$, and hence a significant part of the cooling
process is of a non-equillibrium nature.
Moreover we observe generically a sizable pressure anisotropy at thermalization, which 
is nevertheless entirely understood in terms of dissipative hydrodynamics. 
An effective thermalization time $w^{(th)}=T_{eff}^{(th)} \tau^{(th)}$, i.e. thermalization time 
measured in units of effective temperature at thermalization depends on the initial 
state, but not strongly, and is between 0.37 and 0.67 for all considered initial 
states (a reasonable RHIC estimate $T=500\,MeV$ and $\tau=0.25\,fm/c$ gives $w=0.63$).
Finally, let us note that we could associate with all these initial data, 
an initial entropy already at $\tau=0$ due to the presence of an apparent horizon.
This observation shows that thermalization and applicability of (all-order)
viscous hydrodynamics is not necessarily associated with the sudden 
appearance of a horizon.


\noindent{\bf Acknowledgments:} We thank A. Rostworowski for collaboration 
during the initial stages of the project, W. Florkowski, 
J.-Y. Ollitrault and T. Roma{\'n}czukiewicz for answering many questions. 
This work was supported 
by Polish science funds as research projects N N202 105136 (2009-2012) and 
N N202 173539 (2010-2012), the Foundation for Polish Science and the Foundation for
Fundamental Research on Matter (FOM), part of the Netherlands Organization for Scientific Research (NWO).

\end{document}